\documentclass[sigconf]{acmart}
\usepackage{enumitem}
\AtBeginDocument{%
  }

\setcopyright{rightsretained}
\copyrightyear{2025}
\acmYear{2025}
\acmMonth{09}
\acmDOI{}
\acmConference[Genaiecom '25]{Proceedings of the second workshop on Generative AI for E-Commerce}{September 22, 2025}{Prague, CZ}
\acmBooktitle{Proceedings of the second workshop on Generative AI for E-Commerce 2025, September 22, 2025}
\acmISBN{}
\acmPrice{15.00}

\begin{document}

\title[Knowledge-Augmented Relation Learning for Complementary Recommendation with Large Language Models]{Knowledge-Augmented Relation Learning for\\ Complementary Recommendation with Large Language Models}

\author{Chihiro Yamasaki}
\orcid{0009-0005-4463-3242}
\affiliation{%
  \institution{The University of Electro-Communications}
  \city{Chofu}
  \state{Tokyo}
  \country{Japan}
}
\email{c.yamasaki@librict.jp}

\author{Kai Sugahara}
\orcid{0009-0003-2367-020X}
\affiliation{%
  \institution{The University of Electro-Communications}
  \city{Chofu}
  \state{Tokyo}
  \country{Japan}
}
\email{research@kais.jp}

\author{Kazushi Okamoto}
\orcid{0000-0002-9571-8909}
\affiliation{%
  \institution{The University of Electro-Communications}
  \city{Chofu}
  \state{Tokyo}
  \country{Japan}
}
\email{kazushi@uec.ac.jp}


\begin{abstract}
Complementary recommendations play a crucial role in e-commerce by enhancing user experience through suggestions of compatible items.
Accurate classification of complementary item relationships requires reliable labels, but their creation presents a dilemma.
Behavior-based labels are widely used because they can be easily generated from interaction logs; however, they often contain significant noise and lack reliability.
While function-based labels (FBLs) provide high-quality definitions of complementary relationships by carefully articulating them based on item functions, their reliance on costly manual annotation severely limits a model's ability to generalize to diverse items.
To resolve this trade-off, we propose Knowledge-Augmented Relation Learning (KARL), a framework that strategically fuses active learning with large language models (LLMs).
KARL efficiently expands a high-quality FBL dataset at a low cost by selectively sampling data points that the classifier finds the most difficult and uses the label extension of the LLM.
Our experiments showed that in out-of-distribution (OOD) settings, an unexplored item feature space, KARL improved the baseline accuracy by up to 37\%.
In contrast, in in-distribution (ID) settings, the learned item feature space, the improvement was less than 0.5\%, with prolonged learning could degrade accuracy.
These contrasting results are due to the data diversity driven by KARL's knowledge expansion, suggesting the need for a dynamic sampling strategy that adjusts diversity based on the prediction context (ID or OOD).
\end{abstract}




\maketitle

\section{Introduction}\label{sec-introduction}

The evolution of e-commerce has driven a paradigm shift in recommender systems, moving beyond individual item suggestions toward optimizing user experience through strategic item combinations~\cite{2020jan_d.xu,2024mar_l.li}. 
Complementary recommendation, a key technique in this domain, aims to identify functionally compatible item pairs to enhance user satisfaction and boost sales~\cite{2024mar_l.li,2024oct_z.li}, an impact proven by results such as a 9.56\% increase in the Visit Buy Rate on the online supermarket platform Meituan Maicai~\cite{2023jun_h.chen} and a 0.23\% boost in product sales on Amazon~\cite{2020oct_j.hao}.

One of the most fundamental challenges in providing effective complementary recommendations is defining and creating labels that accurately describe the relationships between items~\cite{2024oct_k.sugahara}.
Early studies in this field relied primarily on behavior-based labels (BBLs)~\cite{2015aug_j.mcauley,2020oct_j.hao}, which are derived from co-occurrence patterns in user interaction data, such as purchase or viewing histories.
Although BBLs can be easily constructed with sufficient user behavioral data and have been widely used to train and evaluate complementary recommendation models, they often suffer from a lack of interpretability and potential noise owing to the inconsistencies or biases inherent in user behaviors~\cite{2023sep_r.papso,2024oct_z.li,2024oct_k.sugahara}.
Consequently, Sugahara et al. proposed function-based labels (FBLs)~\cite{2024oct_k.sugahara,2025jul_c.yamasaki}.
Rather than merely indicating the presence or absence of a complementary relationship, FBLs classifies item pairs into nine functional categories.
These labels are annotated by domain experts, and are completely independent of user interaction logs, making them reliable and noise-resistant alternatives to BBLs.

A practical drawback of FBLs is that they are not data-driven, resulting in substantial annotation costs.
On e-commerce platforms with a large number of items, annotating all possible item combinations using FBLs is practically impossible.
A more feasible approach would be to allocate human resources to annotate a subset of items concentrated in specific item categories on an e-commerce site.
Recommendation models trained on such limited datasets perform well within the in-distribution (ID) feature space~\cite{2025jul_c.yamasaki}; however, their generalization performance for out-of-distribution (OOD) items is often disappointing.
From a different perspective, recent advancements in annotation techniques have demonstrated that large language models (LLMs) can function as zero-shot classifiers~\cite{2024oct_z.li,2025jul_c.yamasaki}, providing an alternative annotator for FBLs.
Nevertheless, the inference cost of LLMs remains impractical for large-scale applications, and LLMs have inherent limitations in covering the entire range of items.

In practice, only a small fraction of item pairs exhibit complementary or substitutable relationships, making it potentially sufficient to efficiently collect informative samples for training classification models.
In other words, blindly annotating all possible item pairs with FBLs is inefficient for learning accurate decision boundaries.
Actively selecting informative item pairs that help the model learn such boundaries, particularly for recognizing complementary relationships, can significantly reduce the need for exhaustive annotations.
Given that the annotation cost, whether by human annotators or large language models (LLMs), depends on the selected item pairs, this active learning approach~\cite{2009_b.settles,2023jan_a.tharwat} enables the construction of effective complementary recommendation models at a minimal scale while still covering diverse item categories.

\begin{figure}[t!]
  \centering
  \includegraphics[width=\linewidth]{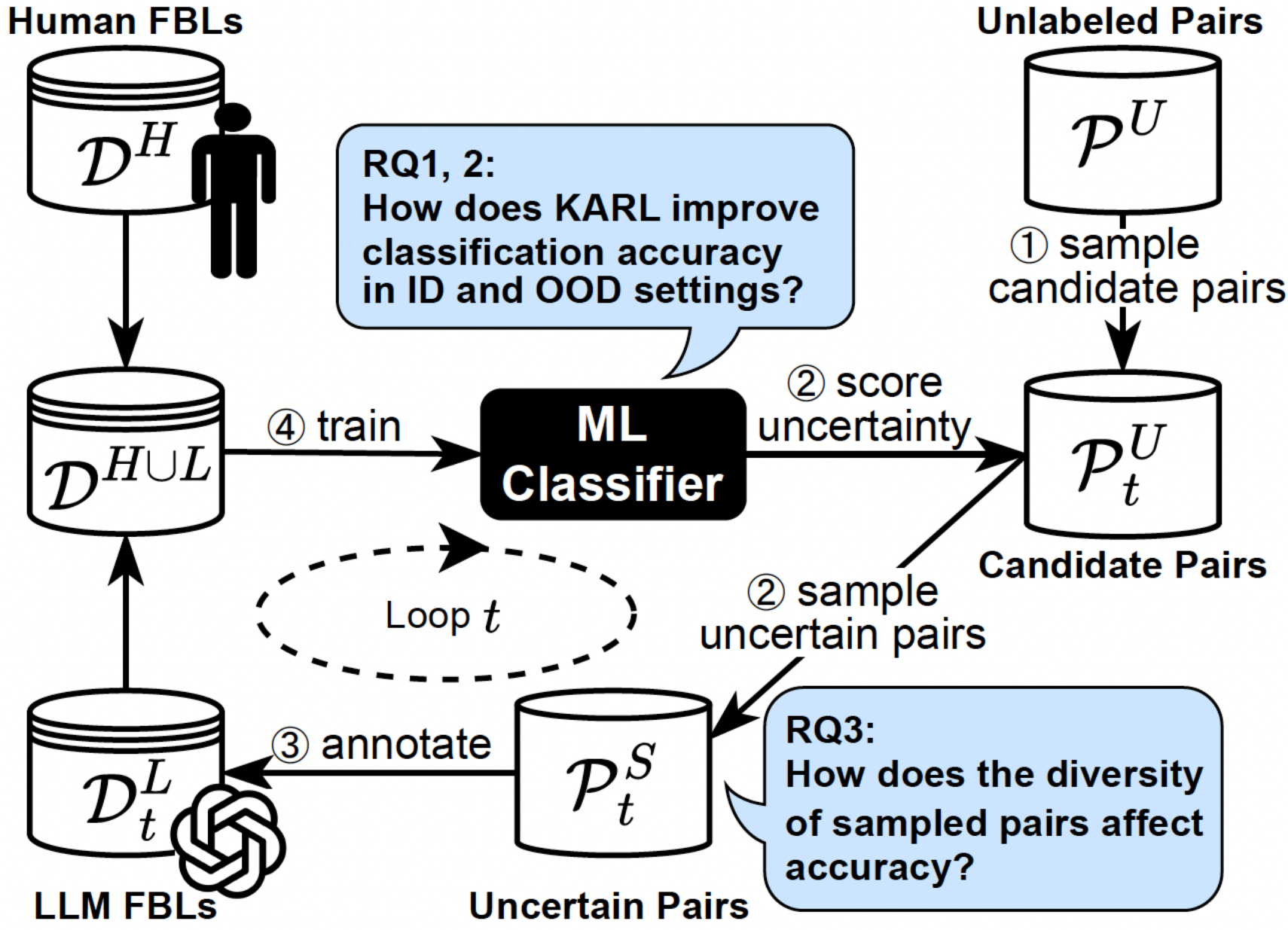}
  \caption{General overview of our study.} \label{fig:overview}
\end{figure}

Drawing from these perspectives, we propose KARL (Knowledge-Augmented Relation Learning), a framework designed to accurately and effectively classify complementary relationships under constrained annotation resources and limited training datasets regarding FBLs.
KARL leverages active learning to iteratively (1) sample uncertain pairs from unlabeled item pairs for where the machine learning (ML) classifier struggles to assign FBLs, (2) assign FBLs to these pairs using LLMs, and (3) retrain the classifier on augmented data, as illustrated in Figure~\ref{fig:overview}.
This iterative process progressively improves the generalization capability of the model by gradually incorporating LLM-generated knowledge, addressing the risk that models trained on small, domain-limited datasets may only perform well in ID settings.
To evaluate its effectiveness, we validated the accuracy of the classifier retrained iteratively through KARL on two FBL datasets corresponding to the ID and OOD scenarios with real-world data from a large e-commerce platform.
Then, we analyzed the effect of the diversity of the training data induced by active learning on the classification accuracy in each scenario.
This study is guided by the following research questions:
\begin{description}[topsep=2pt, partopsep=0pt, itemsep=2pt, parsep=0pt]
  \item{{\textbf{RQ1:}} To what extent does KARL enhance the generalization accuracy in ID feature spaces compared with the baseline?}
  \item{{\textbf{RQ2:}} To what extent does KARL enhance the generalization accuracy in OOD feature spaces compared with the baseline?}
  \item{{\textbf{RQ3:}} How does training data diversity driven by uncertainty sampling affect classification accuracy in ID and OOD settings?}
\end{description}
\section{Related Work}\label{sec-relatedWork}

\subsubsection*{\textbf{Complementary Recommendation.}}
With the growth of e-commerce, complementary recommendations have become the key to enhancing user experience and increasing sales~\cite{2021dec_n.entezari}. 
Previous studies in this field primarily utilized BBLs~\cite{2015aug_j.mcauley,2020oct_j.hao}, derived from user interaction logs such as purchasing or viewing histories. 
Although BBLs can be constructed at a low cost on a large scale and have been widely used for training and evaluating complementary recommender systems~\cite{2018feb_z.wang,2018sep_y.zhang,2021dec_l.ma,2023apr_k.bibas,2023sep_r.papso,2024feb_h.luo}, their reliability is often compromised by noise labels stemming from 
irregular user behaviors~\cite{2020jan_d.xu,2021dec_l.ma,2023sep_r.papso,2024oct_z.li,2024oct_k.sugahara}.
To address the challenges of BBLs, FBLs~\cite{2024oct_k.sugahara,2025jul_c.yamasaki} have been proposed to define item relationships based on the following nine functional categories:
\begin{enumerate}[topsep=2pt, partopsep=0pt, itemsep=2pt, parsep=0pt]
    \item[(A)] Items $x$ and $y$ have the same function and usage.
    \item[(B-1)] Item $x$ can be replenished with item $y$.
    \item[(B-2)] Item $y$ can be replenished with item $x$.
    \item[(C-1)] Items $x$ and $y$ must be combined to be usable.
    \item[(C-2)] When combined with item $y$, item $x$ becomes more useful.
    \item[(C-3)] When combined with item $x$, item $y$ becomes more useful.
    \item[(C-4)] Combining $x$ and $y$ makes them more useful.
    \item[(D)] Items $x$ and $y$ have no relationship.
    \item[(E)] Items $x$ and $y$ seem to have a relationship, but it is difficult to verbalize.
\end{enumerate}
FBLs are highly reliable because they are defined by domain experts based on functional relationships and are independent of noisy user behaviors, leading to high classification accuracy when used for training~\cite{2025jul_c.yamasaki}. 
However, this reliance on manual expert annotation, although integral to their quality, makes expanding FBL datasets prohibitively expensive.

To address the high cost of FBLs, a recent work demonstrated that LLMs render effective annotators for FBL creation~\cite{2025jul_c.yamasaki}.
Specifically, GPT-4o-mini~\cite{2024oct_openai} achieved a macro-F1 score of 0.849 with human ground truth in 3-class classification (complementary, substitute, and unrelated).
Meanwhile, similar studies applying LLMs to relationship classification still face unresolved challenges. 
Although a study using an LLM directly as a classifier reported high accuracy and explainability for the identified relationships~\cite{2024oct_z.li}, this approach faces scalability issues when applied to numerous item combinations in e-commerce. 

\subsubsection*{\textbf{LLM-based Active Learning.}}
Active learning is a technique for efficiently training models while minimizing annotation costs~\cite{2009_b.settles}.
This works by having the model actively select samples that are difficult to predict, which are then labeled by a domain expert and added to the training set.~\cite{2022aug_e.mosqueira,2023jan_a.tharwat}. 
Common sampling strategies include uncertainty sampling, which selects samples near the decision of the model boundary~\cite{2009_b.settles,2022aug_e.mosqueira,2023jan_a.tharwat,2022sep_a.hein}, and diversity sampling, which prevents data bias~\cite{2003aug_k.brinker,2023jan_a.tharwat,2025may_y.xia}. 
Traditionally, active learning has relied on human annotators; this is a highly reliable approach, but remains costly~\cite{2023jan_a.tharwat,2024sep_n.kholodna}.

In response, the remarkable inference capabilities of LLMs have shown promise as cheaper alternatives to human annotators~\cite{2025may_y.xia,2025apr_c.tseng,2024sep_n.kholodna,2023dec_r.xiao}. 
For instance, using GPT-4-Turbo for low-resource language annotation has been reported to significantly reduce costs compared with human annotation~\cite{2024sep_n.kholodna}. 
In addition, the benefits may extend beyond cost reduction, as some studies have reported a synergistic effect in which a model trained on LLM-generated labels surpasses the performance of the LLM itself~\cite{2023nov_r.zhang,2024oct_y.cui}. 
However, the promising performance of LLMs is not universal. 
A comprehensive experiment confirmed that their effectiveness is highly dependent on the specific task and data characteristics~\cite{2023may_n.pangakis}. 
This highlights the critical need to validate the applicability and limitations of the LLM-based annotations in specific domains.
\section{Methodology}\label{sec-framework}

We propose KARL, an active learning framework designed to achieve an accurate and effective classification of item relationships by efficiently leveraging limited FBL datasets and annotation resources.
The framework utilizes the following two data sources: $\mathcal{P}^{U}$, which consists of all possible unlabeled item pairs, and $\mathcal{D}^{H}$, a dataset of human-annotated FBLs used to train the classifier.
The framework initially trains a classifier on $\mathcal{D}^{H}$ for 3-class classification: \textit{complementary}, \textit{substitute}, and \textit{unrelated}.
Following the methodology and the prompt from previous studies~\cite{2025jul_c.yamasaki}, we used Bayesian-optimized~\cite{2019_t.akiba} logistic regression with a 424-dimensional content-based feature vector, and performed LLM-based annotation using GPT-4o-mini~\cite{2024oct_openai}.
Once initialized, KARL enhances the classifier through the following iterative four-step process, which is illustrated in Figure~\ref{fig:overview}:

\begin{description}[topsep=2pt, partopsep=0pt, itemsep=2pt, parsep=0pt]

\item{\textbf{Step 1: Candidate Pair Sampling}}
Because processing the entire set of all possible pairs $\mathcal{P}^{U}$ is computationally infeasible owing to memory constraints, we sample a subset $\mathcal{P}^{U}_{t}$ in each round $t$.
We employed a two-stage hierarchical sampling approach to ensure that this subset was not biased towards specific categories.
We selected up to 10 query items from each of the 368 fine-grained categories (e.g., ballpoint pens), and thereafter paired each with up to 100 candidates from the same broad category (e.g., Office Supplies).

\item{\textbf{Step 2: Uncertainty Sampling}}
The uncertainty scores were calculated based on the model prediction probabilities for each pair in $\mathcal{P}^{U}_{t}$ and the most uncertain pair per fine-grained category was selected as $\mathcal{P}^{S}_{t}$.
This ensures diverse relationship representation while preventing similar-pair dominance.

\item{\textbf{Step 3: LLM-Based Annotation}}
The LLM annotates each pair from $\mathcal{P}^{S}_{t}$ into nine FBLs classes using a prompt that incorporates the description of the pair.
The 9-class output is then systematically mapped to the traditional 3-class: FBLs(A) map to \textit{substitute}, FBLs(B-*,C-*) map to \textit{complementary}, and FBLs(D,E) map to \textit{unrelated}.
We applied a consistency protocol to ensure the reliability of the LLM-annotated labels~\cite{2023may_n.pangakis}. 
Under this protocol, only pairs in which the three independent labels are identical are adopted into $\mathcal{D}^{L}_{t}$.

\item{\textbf{Step 4: Model Retraining}}
The accumulated $\mathcal{D}^{L}$ is integrated with $\mathcal{D}^{H}$ for retraining.
We employed bagging ensemble to address the class imbalance in $\mathcal{D}^{L}$~\cite{1994sep_l.breiman}: ten balanced subsets from $\mathcal{D}^{L}$ via random undersampling were each combined with $\mathcal{D}^{H}$ to train separate classifiers.
The final predictions were averaged across the classifiers, with the aggregated probabilities used in Step 2 of the next round.

\end{description}
\section{Experiments}\label{sec-experiments}

\subsection{Experimental Setup}

We observed changes in the ML classifier accuracy and training data diversity over 20 active learning loop rounds to evaluate the efficiency and convergence properties of KARL.

\subsubsection*{\textbf{Datasets.}} 
Our study utilized an item dataset provided by ASKUL Corporation\footnote{https://www.askul.co.jp/corp/english/} including rich item attributes such as title, description, hierarchical categories, and more.
To test how well KARL works in ID and OOD settings, we used two human-annotated FBLs datasets, $\mathfrak{D}^{H}_{id}$ and $\mathfrak{D}^{H}_{ood}$\footnote{\url{https://github.com/okamoto-lab/fbl_dataset}}~\cite{2024oct_k.sugahara,2025jul_c.yamasaki}:
\begin{itemize}
  \item $\mathfrak{D}^{H}_{id}$ (ID): This set contains 2,625 labeled pairs, sampled based on high co-occurrence patterns and supplemented with web-sourced pairs.
  This dataset comprises 591 complementary, 410 substitute, and 1,624 unrelated pairs.
  \item $\mathfrak{D}^{H}_{ood}$ (OOD): This set contains 2,790 labeled pairs created by sampling one query item from each of 366 fine-grained categories and pairing it with another item based on BBLs. 
  This dataset comprises 375 complementary, 2,024 substitute, and 391 unrelated pairs.
\end{itemize}
Both comprised item pairs from the ``Office Supplies/Stationery'' and ``Household Goods/Kitchenware'' categories, with three independent human annotations per pair, from which we retained only pairs with majority-agreed labels.
To evaluate ID accuracy, we employed 5-fold nested cross-validation~\cite{2009feb_p.filzmoser} on $\mathfrak{D}^{H}_{id}$, training models on each training fold and reporting averaged results across all test folds.
For OOD evaluation, we tested these models on the entire $\mathfrak{D}^{H}_{ood}$.
The severe distributional shift between these datasets is evident: a classifier trained on $\mathfrak{D}^{H}_{id}$ achieved only 0.44 macro-F1 on $\mathfrak{D}^{H}_{ood}$, as detailed in Section 4.3.
In addition, our unlabeled pair pool $\mathcal{P}^{U}$ consisted of item pairs within the same item categories found on the same e-commerce source as the FBL datasets.

\subsubsection*{\textbf{Uncertainty Sampling Methods.}}
We compared three uncertainty sampling methods to score uncertainty in Step 2:
\begin{enumerate}
  \item \textbf{Random}: Baseline method for randomly selecting pairs by assigning uniform random scores to each pair.
  \item \textbf{Query-by-Committee (QBC)}~\cite{1992jul_h.s.seung}: Select pairs with the highest prediction variance across ten bagging models, where variance is computed over the predicted class probabilities.
  \item \textbf{Margin}~\cite{2001_t.scheffer,2002mar_s.tong}: Select pairs with the smallest probability margin between the top two predicted classes.
\end{enumerate}

\subsubsection*{\textbf{Evaluation Metrics.}}
We evaluated KARL using two metrics: classification accuracy and training data diversity.
Classification accuracy was measured by macro-F1, which averages the F1-score across all classes. 
Training data diversity was quantified using a Pearson correlation metric $\rho: \mathbb{R}^{d} \times \mathbb{R}^{d} \longrightarrow [-1, 1]$:
$$
diversity(X) = 1 - \frac{1}{n(n - 1)} \sum_{i \ne j} \left| \rho(X_{i \cdot}, X_{j \cdot}) \right|
$$
where $n$ is the number of training pairs and $X_{i \cdot}$ denotes the feature vector of the $i$-th pair.
Higher $diversity(X)$ show greater diversity.
While other diversity metrics might provide complementary insights, we adopted this correlation-based approach as a simple baseline for quantifying diversity.

\subsection{ID Accuracy Analysis (RQ1)}
\begin{figure}[t] 
  \centering 
  \includegraphics[width=\linewidth]{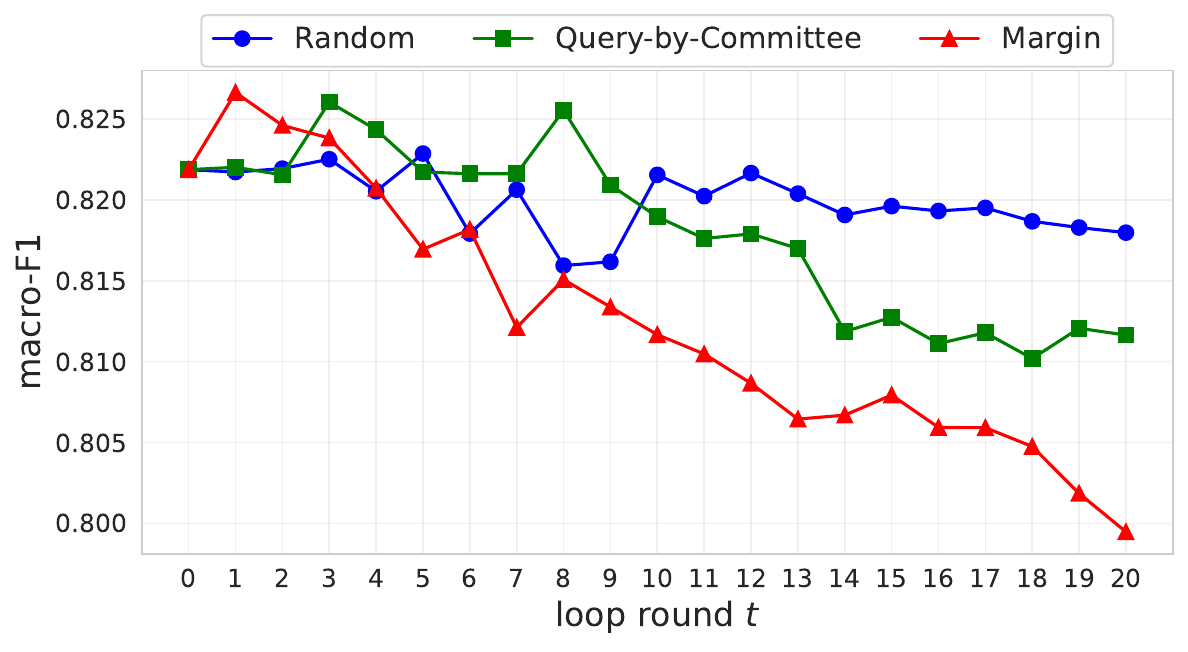} 
  \caption{ID macro-F1 score on $\mathfrak{D}^{H}_{id}$, averaged across the 5 test folds of nested cross-validation.} \label{fig:macro-f1-id} 
\end{figure}

Figure~\ref{fig:macro-f1-id} shows the macro-F1 progression on the ID test set over 20 rounds. 
Whereas the baseline model (loop 0) started with high accuracy, consistent with previous studies~\cite{2025jul_c.yamasaki}, KARL offered only marginal gains ($\leq$ 0.5\%) before steadily degrading the accuracy. 
This degradation was most pronounced with uncertainty-based sampling methods such as QBC and Margin.
This result suggests that the model has already captured sufficient ID relationships, making additional diverse data counterproductive. 
Rather than providing useful information, ambiguous samples from uncertainty sampling acted as ``noise'' that disrupted the model's stable distribution of the model and caused an accuracy-degrading shift. 

\subsection{OOD Accuracy Analysis (RQ2)}
\begin{figure}[t] 
  \centering 
  \includegraphics[width=\linewidth]{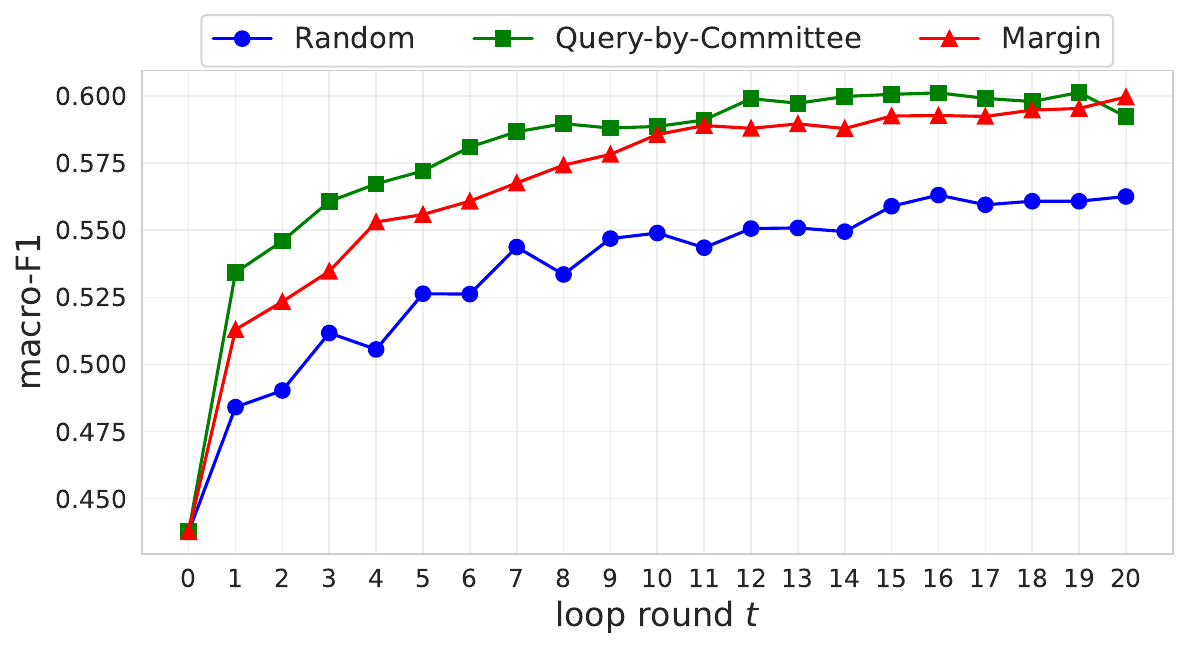} 
  \caption{OOD macro-F1 score on $\mathfrak{D}^{H}_{ood}$.} \label{fig:macro-f1-ood} 
\end{figure}

In contrast to the ID scenario, KARL proved highly effective for the OOD test set.
As shown in Figure~\ref{fig:macro-f1-ood}, KARL dramatically improved macro-F1 by up to 37\% compared to the baseline.
The superiority of the uncertainty sampling methods (QBC, Margin) over Random is particularly noteworthy, offering two distinct advantages.
First, they are significantly more cost-efficient; they achieve any given level of accuracy in far fewer rounds than Random, thus minimizing additional annotation costs. 
Second, they increase the peak accuracy of the model. 
Although all the methods eventually plateaued at around loop 15, the final accuracy achieved by the uncertainty-based methods was up to 6.6\% higher than that of Random, resulting in a more capable classifier.
This evidence demonstrates that selective sampling in an OOD context not only accelerates learning but also raises the ceiling of the potential of the model.

\subsection{Diversity--Accuracy Relation Analysis (RQ3)}

\begin{figure}[t] 
  \centering 
  \includegraphics[width=\linewidth]{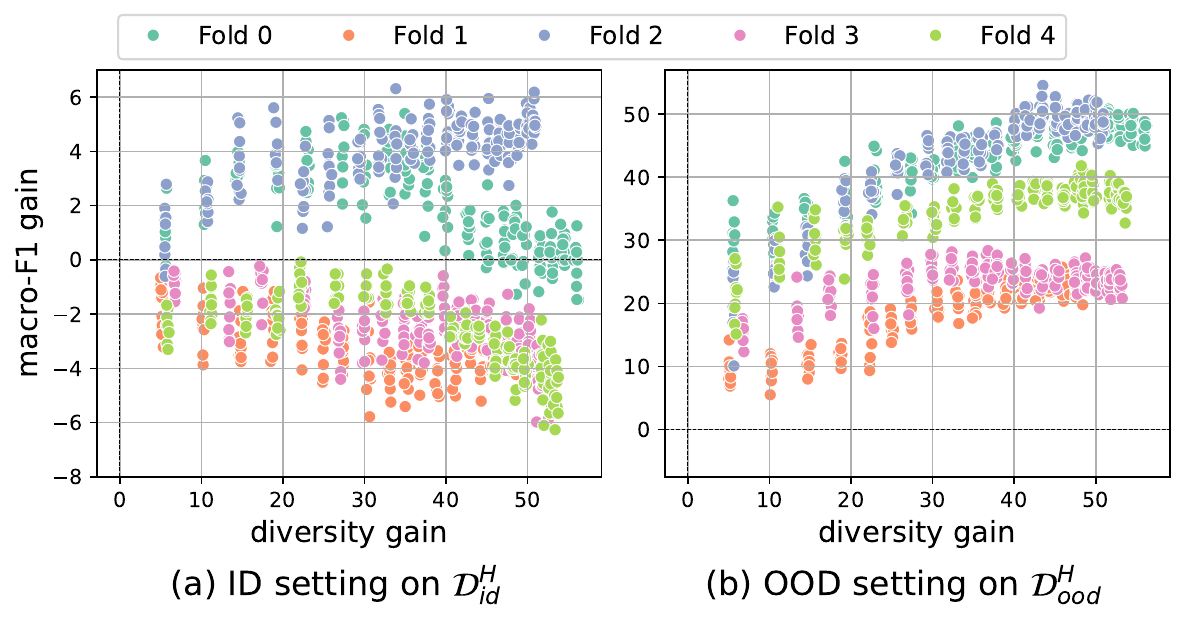} 
  \caption{Correlation between accuracy gain and diversity gain in the training set for QBC.} \label{fig:gain-qbc} 
\end{figure}

\begin{figure}[t] 
  \centering 
  \includegraphics[width=\linewidth]{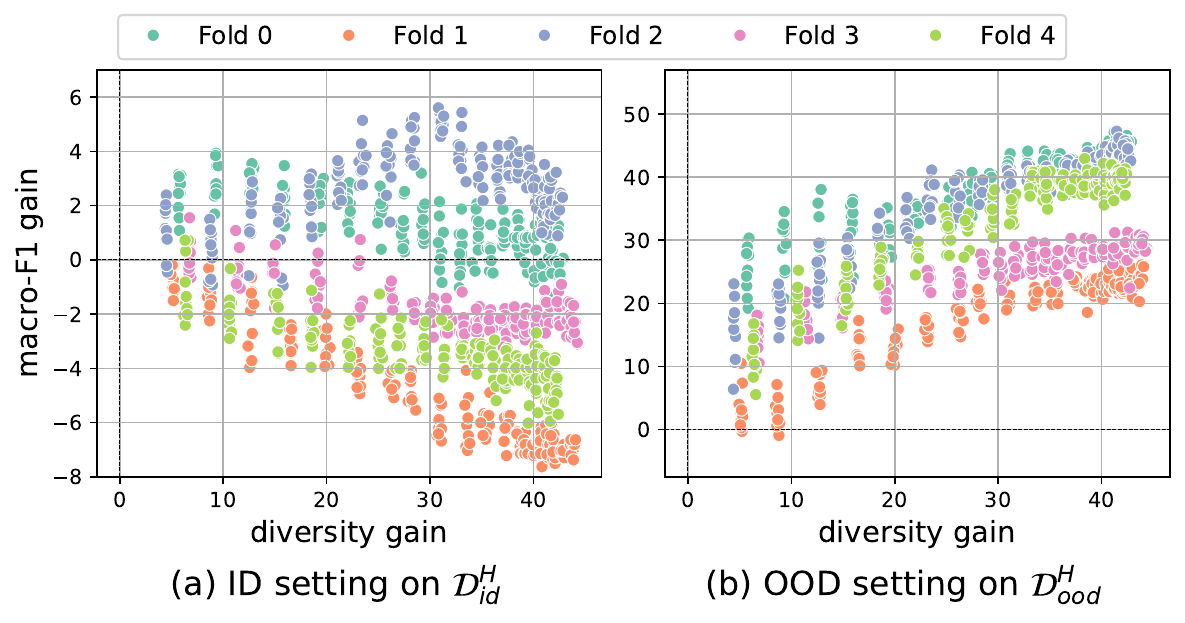} 
  \caption{Correlation between accuracy gain and diversity gain in the training set for Margin.} \label{fig:gain-margin} 
\end{figure}

To understand the mechanisms behind the contrasting results in RQ1 and RQ2, we analyzed the correlation between training data diversity and classifier accuracy gains.
In Figures~\ref{fig:gain-qbc} and~\ref{fig:gain-margin}, we plot the accuracy gain against the diversity gain, with both metrics measured as changes from the baseline at loop 0.
Each scatter plot shows the correspondence for a specific model at a specific loop round, comprising 200 data points--derived from the product of outer folds, loop rounds, and bagging models.

In the ID setting (panel (a)), although some folds showed a modest initial accuracy improvement, adding diversity beyond a certain threshold proved counterproductive, leading to a consistent decline in accuracy.
Conversely, in the OOD setting (panel (b)), the diversity gain exhibits a strong and consistent positive correlation with the macro-F1 gain across all folds, resulting in a substantial improvement in accuracy of up to approximately 50\%.
This demonstrates that while diversity is a key driver of knowledge expansion in unfamiliar feature spaces, it can disrupt well-learned distributions in familiar ones.

\section{Conclusion}\label{sec-conlusion}

This study presented KARL, a framework that addresses the cost and scalability challenges of FBLs by synergizing a cost-effective LLM annotator with an active learning strategy prioritizing informative samples.
Our experiments demonstrated the high effectiveness of KARL in OOD settings, where increased data diversity directly promoted the acquisition of new knowledge, leading to improved generalization.
Conversely, its effectiveness was limited in ID settings, as the excessive pursuit of diversity proved counterproductive to knowledge refinement by disrupting the learned data distribution.
These contrasting results suggest that future frameworks should implement context-aware dual modes: preserving learned distributions in ID settings while aggressively exploring in OOD settings.
Such adaptive strategies could use confidence thresholds to automatically switch between conservative and exploratory sampling.

This study has several limitations that point to future research.
First, the use of logistic regression may limit accuracy in OOD settings, as its linear decision boundaries may be insufficient for capturing complex non-linear complementary relationships.
Future work should explore non-linear models such as gradient boosting or neural networks.
Second, while our framework relies on GPT-4o-mini for annotation, the generalizability to other LLMs remains unexplored.
Future studies should compare different LLMs and develop better methods for estimating label quality.

\begin{acks}
This work was supported by ASKUL Corporation and JSPS KAKENHI Grant Numbers JP23K21724, JP24K21410.
\end{acks}

\bibliographystyle{ACM-Reference-Format}
\bibliography{references.bib}


\end{document}